\documentclass[accept]{jfm}

\usepackage{graphicx}
\usepackage{epstopdf,epsfig}
\usepackage{newtxtext}
\usepackage{newtxmath}
\usepackage{natbib}
\usepackage{hyperref}
\hypersetup{
    colorlinks = true,
    urlcolor   = blue,
    citecolor  = blue,
}

\newcommand{\RomanNumeralCaps}[1]
\linenumbers
\newcommand{\uFluid}[0]{\boldsymbol{u}}

\newcommand{\StressTensor}[0]{{\mathsfbi{S}}}
\newcommand{\HeatFlux}[0]{{\boldsymbol{Q}}}

\newcommand\Ma{\mbox{\textit{Ma}}}

\usepackage{soul} 

\newcommand{\commentout}[1]{}

\title{Molecular Fluctuations Inhibit Intermittency in Compressible Turbulence}

\author{Ishan Srivastava\aff{1}
  \corresp{\email{isriva@lbl.gov}},
  Andrew J. Nonaka\aff{1},
  Weiqun Zhang\aff{1},
  Alejandro L. Garcia\aff{2},
 \and John B. Bell\aff{1}}

\affiliation{\aff{1}Center for Computational Sciences and Engineering, Lawrence Berkeley National Laboratory, 1 Cyclotron Road, Berkeley, California 94720, USA
\aff{2}Department of Physics and Astronomy, San Jose State University, 1 Washington Square, San Jose, California 95192, USA}

\begin{document}
\maketitle

\begin{abstract}
In the standard picture of fully-developed turbulence, highly intermittent hydrodynamic fields are nonlinearly coupled across scales, where local energy cascades from large scales into dissipative vortices and large density gradients. Microscopically, however, constituent fluid molecules are in constant thermal (Brownian) motion, but the role of molecular fluctuations on large-scale turbulence is largely unknown, and with rare exceptions, it has historically been considered irrelevant at scales larger than the molecular mean free path. Recent theoretical and computational investigations have shown that molecular fluctuations can impact energy cascade at Kolmogorov length scales. Here we show that molecular fluctuations not only modify energy spectrum at wavelengths larger than the Kolmogorov length in compressible turbulence, but they also significantly inhibit spatio-temporal intermittency across the entire dissipation range. Using large-scale direct numerical simulations of computational fluctuating hydrodynamics, we demonstrate that the extreme intermittency characteristic of turbulence models is replaced by nearly-Gaussian statistics in the dissipation range. These results demonstrate that the compressible Navier-Stokes equations should be augmented with molecular fluctuations to accurately predict turbulence statistics across the dissipation range. Our findings have significant consequences for turbulence modeling in applications such as astrophysics, reactive flows, and hypersonic aerodynamics, where dissipation-range turbulence is approximated by closure models. 
\end{abstract}

\begin{keywords}
\end{keywords}


\section{Introduction}
\label{sec:intro}
A fully developed three-dimensional turbulent state is highly irregular with energy nonlinearly `cascading' from large length scales where it is injected to small length scales in an essentially inviscid process, until it is eventually dissipated by the viscosity of the fluid at scales smaller than the dissipation length scale (also known as the Kolmogorov length scale) \citep{Frisch1995,alexakis2018cascades,eyink2006onsager}. In incompressible fluids, the energy cascades occurs by a continuous transition of large eddies into smaller and smaller eddies while energy is continually injected at large length scales in a nonequilibrium statistical steady state. Such a cascading phenomenon indicates that the statistical properties of turbulence should be invariant at all scales, as predicted by Kolmogorov's theory of turbulence \citep{Frisch1995}. However, intermittency in turbulent flows result in strong deviations from Kolmogorov's theory at small scales \citep{frisch1981intermittency,Frisch1995,paladin1987anomalous,chevillard2005rapid}. Intermittency is characterized by extreme variability of velocities with non-Gaussian, fat-tailed distributions that appear as localized bursts of extreme vorticity intensification in a largely quiescent flow \citep{yeung2015extreme,benzi2008intermittency,wang2017scaling}.

While energy cascades and intermittency have been intensely studied in incompressible fluids, numerous natural and technological phenomena involve turbulent flow of compressible fluids. Important natural applications include astrophysical phenomena such as supernovae, star formation and cosmology \citep{mac2004control}. Compressible turbulence is also important in technological applications such as high-temperature reactive flows \citep{hamlington2012intermittency}, inertial confinement fusion \citep{bender2021simulation}, and hypersonic vehicle design \citep{urzay2018supersonic}. The dynamics of compressible turbulence is significantly more complicated than incompressible turbulence with nonlinear interactions between solenoidal (shear) and compressive modes of velocity fluctuations along with coupling between the velocity field and thermodynamic fields (pressure and density) \citep{eyink2018cascades}. For example, in addition to dissipative vortices, compressible turbulence is also characterized by the appearance of shock waves \citep{federrath2021sonic} and contact surfaces characterized by large density gradients \citep{benzi2008intermittency}. Whereas exact scaling relations for the correlation functions and statistical properties of compressible turbulence have been recently discovered \citep{wang2012scaling,donzis2020universality,eyink2018cascades,wang2017scaling}, further analysis suggests that kinetic energy dissipation occurs due to a distinct mechanism of pressure-work defect \citep{eyink2018cascades} in addition to local energy cascades \citep{aluie2011compressible,wang2013cascade}. However, despite more complex physical mechanisms, turbulent compressible flows also exhibit local energy cascades, which minimally conserve kinetic energy \citep{aluie2011compressible,wang2013cascade}, and strongly intermittent and variable velocity and thermodynamic fields at smaller length scales \citep{benzi2008intermittency,wang2017scaling,federrath2021sonic}.

Microscopically, fluids are a discrete physical system consisting of molecules that are in constant random (i.e., Brownian) motion; an accurate continuum description at small scales requires the use of fluctuating fields. Unlike turbulent fluctuations described above, these molecular fluctuations are thermal in origin with a covariance structure that is completely described by equilibrium statistical mechanics \citep{landau_statistical_1980}.  While thermal fluctuations are present at all scales in a fluid, in nonequilibrium conditions fluctuations in velocity and thermodynamic fields can become correlated over macroscopic length scales, resulting in interesting macroscale phenomena such as non-equilibrium correlations observed in light scattering \citep{tremblay1981fluctuations}, diffusive enhancement by mode coupling \citep{DonevPRL2011}, giant fluctuations \citep{GiantFluctuations_Nature}, and hydrodynamic instabilities \citep{RayleighBernard_Fluctuations}. It is therefore an important question to ask: at what scales do thermal fluctuations have a significant effect on turbulent fluctuations?  While it has been historically accepted that thermal fluctuations do not impact turbulence at scales larger than the mean free path \citep{neumann1949recent}, recent \citep{bandak2022dissipation} and rediscovered \citep{Betchov1957} theoretical efforts have remarkably predicted that thermal fluctuations can dominate the kinetic energy spectrum at scales comparable to the dissipative Kolmogorov length scale which is orders of magnitude larger than the mean free path of most common fluids. These theoretical predictions have been confirmed by very recent modeling efforts \citep{Bell2022thermal,mcmullen2022navier}, but no experimental confirmation exists. While a recent numerical study on incompressible fluids study has discovered that molecular fluctuations replace the extreme-scale intermittency in the far-dissipation range with a Gaussian distribution \citep{Bell2022thermal}, the impact of molecular fluctuations on turbulent intermittency across the whole range of turbulence spectrum remains to be determined. Furthermore, the impact of molecular fluctuations on compressible turbulence has also not been fully explored.

\section{Theory and numerical methods}
\label{sec:theory_numeric}
\subsection{Fluctuating hydrodynamics theory of compressible fluids}
In order to reliably introduce thermal fluctuations in compressible fluid dynamics, we use nonlinear \emph{fluctuating hydrodynamics} (FHD), originally proposed in the linearized form by \citet{Landau1959Fluid} (see also \citep{ZarateBook2006}). Here, a stochastic flux term is added to the deterministic Navier-Stokes equations, leading formally to a system of stochastic partial differential equations (SPDEs). The stochastic fluxes represent a macroscopic realization of microscopic degrees of freedom in a thermodynamic system. Specifically, these fluxes are constructed to model fluctuations in hydrodynamic variables that arise from the discrete molecular character of fluids as predicted by statistical mechanics. The linearized form of FHD was justified by \citet{LLNS_FD_Fox,Boltzmann_FD_Fox}, and \citet{PhysRev.187.267}. The nonlinear hydrodynamic fluctuations were later justified by deriving the Fokker-Planck equations of the distribution function of conserved hydrodynamic quantities \citep{zubarev1983statistical}, which then led to the formulation of the associated stochastic differential equations \citep{LLNS_Espanol}.

The nonlinear FHD equations for a compressible fluid in conservative form are \citep{srivastava2023staggered}:
\begin{subeqnarray}
    &\frac{\partial}{\partial t} \left( \rho \right) = - \bnabla\bcdot\left( \rho \uFluid \right),\\[3pt]
    &\frac{\partial }{\partial t} \left( \rho \uFluid \right) =  - \bnabla \bcdot \left[ \rho {\uFluid \otimes \uFluid + p\mathbb{I} } \right] - \bnabla \bcdot \left[ \StressTensor + \widetilde{\StressTensor} \right] + \rho \boldsymbol{a}^{F},\\[3pt]
    &\frac{\partial }{\partial t} \left( \rho E \right) = - \bnabla \bcdot \left[\uFluid \left(\rho E + p\right) \right] - \bnabla \bcdot \left[ \HeatFlux + \widetilde{\HeatFlux} \right] -\bnabla \bcdot \left[ \left( \StressTensor + \widetilde{\StressTensor} \right) \bcdot  \uFluid \right] \\ \nonumber
    &+ \rho\boldsymbol{a}^{F} \bcdot \uFluid - \langle \rho \boldsymbol{a}^{F} \bcdot \uFluid \rangle
    \label{eq1}
\end{subeqnarray}
where $\rho$ is the fluid density, $\uFluid$ is the velocity, $E$ is the total specific energy, $p$ is the pressure, and $\mathbb{I}$ is the identity matrix. The total energy density of the fluid $\rho E = \rho e + \frac12 \rho (\uFluid \bcdot \uFluid)$ is the sum of internal energy and kinetic energy, where $e$ is the specific internal energy. In this set of nonlinear FHD equations, the diffusive stress tensor $\StressTensor$ and heat flux $\HeatFlux$ are augmented by their stochastic counterparts $\widetilde{\StressTensor}$ and $\widetilde{\HeatFlux}$ respectively. When $\widetilde{\StressTensor}=\widetilde{\HeatFlux}=0$, the FHD equations reduce to the well-known \emph{deterministic} Navier-Stokes equations for compressible fluids. The term $\boldsymbol{a}^{F}$ represents a long-wavelength external turbulence acceleration required for maintaining a statistically-steady turbulent state. The last term in the energy equation $- \langle \rho \boldsymbol{a}^{F} \bcdot \uFluid \rangle$ represents a thermostat that is used to maintain the system temperature. The details of the diffusive and stochastic fluxes, and the turbulence forcing and thermostat are presented in the next sections.

The linearized form of FHD equations is a well-defined system of SPDEs with equilibrium solutions that are Gaussian random fields with a covariance structure that matches the Gibbs-Boltzmann distribution that is consistent with well-established results in statistical mechanics \citep{landau_statistical_1980}. Although the linearized FHD equations can be rigorously defined with the use of generalized functions, the high irregularity of the stochastic fluxes makes interpreting the fully nonlinear system as SPDEs mathematically ill-defined. To obtain a mathematically tractable model, one needs to introduce a high wave-number cutoff that is of the order of several mean free paths. 
In practice, we introduce a cutoff by discretizing the system using a finite-volume discretization with cells that are large enough to have at least $N\ge 50$ molecules per finite-volume cell, resulting in a finite-dimensional system of stochastic differential equations \citep{srivastava2023staggered}. This system of stochastic differential equations models the effect of thermal fluctuations as measured at the grid scale. Setting $N\ge 50$ ensures that variation in hydrodynamic variables are well-approximated by a Gaussian. 
The computational methodology used in this work has been demonstrated to accurately capture the effect of thermal fluctuations in both equilibrium and non-equilibrium settings by comparison against theory and molecular gas dynamics simulations \citep{srivastava2023staggered}. We note that the numerical solution of the FHD equations depends on the specific mesh spacing in the finite volume discretization. This reflects the physical property that the variance of fluctuations in hydrodynamic variables depends on the scale at which they are measured.

As such, there is ample numerical evidence that a finite-volume discretization of FHD equations accurately models nonlinear hydrodynamics fluctuations in various macroscale nonequilibrium phenomena such as giant fluctuations \citep{srivastava2023staggered} and diffusive enhancement \citep{DonevPRL2011}.
While FHD has proven remarkably successful for modeling mesoscale laminar flows with thermal fluctuations, matching theory and experiment, numerical solutions of the FHD equations have only very recently been utilized to model turbulence in incompressible fluids with molecular fluctuations \citep{Bell2022thermal}.
Here we consider application of FHD to compressible turbulence. Specifically, we perform direct numerical simulations of homogeneous isotropic turbulence in nitrogen gas at standard temperature and pressure (STP) subjected to a large wavelength random external solenoidal forcing along with a thermostat to maintain a statistically steady turbulent state. The simulation domain is a periodic cube with sides of approximate length $L\approx 0.2$mm discretized on a $1024^3$ finite-volume grid.  The grid size $\Delta x = 1.956\times10^{-4}\text{mm}$ then sets the small wavelength (high wavenumber) cutoff of the numerical solution to the FHD equations that corresponds to the coarse-graining length of the microscopic fluid dynamics. At STP the mean free path of nitrogen molecules is approximately $70$nm, which is about $3$ times smaller than the grid size corresponding to the high wavenumber cutoff. We also restrict the present study to weakly compressible flows with subsonic turbulent Mach numbers $\Ma_{t} \approx 0.2$ that can exhibit large density variations with contact discontinuities even in the absence of hydrodynamic shocks \citep{benzi2008intermittency}.

\subsection{Numerical details}
Here we present the numerical details for solving the nonlinear FHD equations defined in Eq.(\ref{eq1}).
For the case of nitrogen gas simulated here, we assume an ideal gas equation of state:
\begin{equation}
    p = \frac{\rho k_B T}{m},
\end{equation}
where $T$ is the temperature, $m$ is the molecular mass, and $k_B$ is the Boltzmann's constant. We assume calorically perfect gas at STP with constant specific heats of a classical diatomic gas. The components of the stress tensor $\StressTensor$ defined in its Newtonian form are:
\begin{equation}
S_{ij} = -\eta \left( \frac{\partial u_i}{\partial x_j} + \frac{\partial u_j}{\partial x_i}  \right) - \delta_{ij} \left( ( \kappa - \frac{2}{3} \eta ) {\bf \nabla} \cdot {\uFluid} \right),
\label{eqn:stress}
\end{equation}
where $\delta_{ij}$ is the Kronecker delta, $\eta$ is the shear viscosity, and $\kappa$ is the bulk viscosity. The heat flux $\HeatFlux = -\lambda \nabla T$, where $\lambda$ is the thermal conductivity. The viscosity and thermal conductivity are not treated as constants but depend on the local state of the fluid \citep{Giov1}.

The stochastic stress $\widetilde{\StressTensor}$ is a Gaussian random field with zero ensemble mean $\langle \widetilde{\StressTensor} \rangle = 0$, and we use the following efficient form of $\widetilde{\StressTensor}$, as proposed by \citet{LLNS_Espanol,morozov1984langevin}, in this study:
\begin{equation}
    \widetilde{\StressTensor}(\boldsymbol{r},t) = \sqrt{2k_B T \eta} \widetilde{\mathsfbi{\mathcal{Z}}} +
\left(\sqrt{\frac{k_B \kappa T}{3}} - \frac{\sqrt{2k_B \eta T}}{3} \right ) \mathrm{Tr} ( \widetilde{\mathsfbi{\mathcal{Z}}} ) \mathsfbi{I}.
\label{eq:stochstress}
\end{equation}
Here,
\begin{equation}
    \widetilde{\mathcal{Z}} = \frac{1}{\sqrt{2}}\left(\mathsfbi{\mathcal{Z}}+\mathsfbi{\mathcal{Z}}^{T}\right)
    \label{stoch:Z}
\end{equation}
is a symmetric matrix constructed from an uncorrelated Gaussian tensor field $\mathsfbi{\mathcal{Z}}$ with zero mean and unit variance. The stochastic heat flux $\widetilde{\HeatFlux}$ is;
\begin{equation}
    \widetilde{\HeatFlux} = \sqrt{2k_BT^2\lambda}\mathsfbi{\mathcal{Z}}^{(\HeatFlux)},
\end{equation}
where $\mathcal{Z}^{(\HeatFlux)}$ is an uncorrelated three-dimensional Gaussian vector field with zero mean and unit variance.

A staggered-grid discretization based on the method-of-lines approach is used to spatially discretize the stochastic PDEs of compressible FHD. Here, the conserved scalar variables, $\rho$ and $\rho E$, and primitive scalar variables, $p$ and $T$, are discretized at the centers of a finite-volume cell, whereas the vector variables, conserved momentum density $\rho \uFluid$ and velocity $\uFluid$, are discretized on the normal faces of the grid \citep{srivastava2023staggered}. The resulting stochastic ordinary differential equations (ODEs) are integrated explicitly in time using a low-storage third-order Runge-Kutta (RK3) integrator \citep{Donev2010Camcos,srivastava2023staggered}. The staggered-grid numerical method discretely preserves the fluctuation-dissipation balance \citep{LLNS_Staggered}, which has been confirmed by a correct reproduction of the structure factors of hydrodynamic variables at thermodynamic equilibrium \citep{srivastava2023staggered}.

We emphasize here that even though the nonlinear FHD equations and the deterministic Navier-Stokes equations for compressible fluids appear similar with the exception of the stochastic forcing, they are conceptually completely different in their representation of the underlying hydrodynamic phenomena. 
FHD represents a coarse-graining of the molecular description of a fluid with an underlying assumption that the coarse-graining region has a sufficient number of molecules. The hydrodynamic and thermodynamic fields resulting from the coarse-graining have statistical properties that depend on the scale at which they are measured, and which become increasingly irregular at smaller scales. This scale dependence is not an artifact but rather a consequence of the molecular character of the fluid. For computational purposes, a numerical cutoff is introduced that is at least as large as the scale needed to justify the coarse-graining process \citep{DiscreteLLNS_Espanol}.  In the present method, this numerical cutoff is given by the mesh size of the finite volume discretization 
that effectively acts as a low-pass filter for the coarse-grained molecular fluctuations \citep{eyink2024onsager}, and the accuracy of the FHD description is assessed by renormalization group invariance of the model to this cutoff \cite{LLNS_Renormalization}. 
In this regard, the invariance of the FHD model to renormalization group transformation is conceptually different than the traditional numerical convergence of the solution of deterministic Navier-Stokes equations to an underlying continuum model.

\subsection{Turbulence forcing and thermostat}
A statistically steady homogeneous isotropic turbulent state is achieved by forcing the system with a stochastic process using the formulation of \citet{Eswaran1988CompFluids}. An external force $\rho\boldsymbol{a}^{F}(\boldsymbol{r},t)$ corresponding to a long-wavelength acceleration $\boldsymbol{a}^{F}(\boldsymbol{r},t)$ is added to the momentum equation to drive turbulence. The forcing is applied only on wavevectors $\boldsymbol{k}$ whose wavenumbers lie inside the spherical shell of radius $2\sqrt{2}k_0$, such that $|\boldsymbol{k}| \leq 2\sqrt{2}k_0$, where $k_0 = 2\pi/L$.

Mathematically, consider an Ornstein–Uhlenbeck (OU) process for a complex-valued vector $\boldsymbol{b}(\boldsymbol{n},t)$ as:
\begin{equation}
    \text{d}\boldsymbol{b}(\boldsymbol{n}) = \mathsfbi{A}\boldsymbol{b}(\boldsymbol{n})\text{d}t + \mathsfbi{B}\text{d}\boldsymbol{W},
\end{equation}
where $\boldsymbol{n}=\left(n_x,n_y,n_z\right)$ are integer indices such that $1\leq |\boldsymbol{n}| \leq 2\sqrt{2}$ limits the forcing to long wavelengths, and $\boldsymbol{W}$ is a vector of complex Wiener processes. The matrices in the OU process are:
\begin{equation}
    \mathsfbi{A} = \frac{1}{T_L}\mathsfbi{I}, \qquad \mathsfbi{B} = \sigma \sqrt{\frac{1}{T_L}}\mathsfbi{I},
\end{equation}
where $\mathsfbi{I}$ is the identity matrix. Therefore we have \citep{Gardiner1985book}:
\begin{equation}
    \langle \boldsymbol{b}(\boldsymbol{n},t) \bcdot \boldsymbol{b}^{*}(\boldsymbol{n'},t+s) \rangle = \frac{\sigma^2}{2}\text{e}^{-s/T_L}\delta_{\boldsymbol{n},\boldsymbol{n'}},
\end{equation}
where $\sigma$ and $T_L$ control the amplitude and time scale of external forcing. In compressible turbulence, both solenoidal and dilatational modes can be forced independently; in this study, we focus on solenoidal forcing only. To do so, we apply a projection operator $\mathsfbi{P}$ on $\boldsymbol{b}(\boldsymbol{n},t)$ such that $\tilde{\boldsymbol{b}}(\boldsymbol{n},t) = \mathsfbi{P} \bcdot \boldsymbol{b}(\boldsymbol{n},t)$ is projected onto a plane normal to $\boldsymbol{k} = 2\pi \boldsymbol{n}/L$, where:
\begin{equation}
    \mathsfbi{P} = \left(\mathsfbi{I} - \frac{\boldsymbol{k}\boldsymbol{k}^{T}}{|\boldsymbol{k}|^2}\right).
\end{equation}
The real-space turbulence forcing is then formulated as:
\begin{equation}
    \boldsymbol{a}^{F}(\boldsymbol{r},t) = \Real\left[\sum_{1\leq|\boldsymbol{n}|\leq 2\sqrt{2}} \left(\tilde{\boldsymbol{b}}(\boldsymbol{n}) + \tilde{\boldsymbol{b}}^{*}(\boldsymbol{-n})\right) \text{e}^{i\boldsymbol{k}\bcdot \boldsymbol{r}}\right].
\end{equation}

The external turbulence forcing adds energy to the compressible fluid that dissipates as heat causing an increase in the system temperature. To maintain a statistically steady state, energy is continually removed from the system using a sink. At each time step, we compute the mean power due to the external forcing as $\langle \rho(\boldsymbol{r}) \boldsymbol{a}^{F}(\boldsymbol{r}) \cdot \uFluid(\boldsymbol{r})\rangle$, which is uniformly removed as a sink term in the energy equation.
We note that at thermodynamic equilibrium without forcing in FHD simulations, no sink is needed because the fluctuation-dissipation balance ensures a statistically-steady state.

\subsection{Simulation details and statistics}

We ran simulations with the initial state of nitrogen gas at STP conditions of density $\rho_0=1.13\times10^{-3}\text{g}/\text{cm}^3$ and $T=300\text{K}$, where the mean free path of nitrogen molecules is $\approx70\text{nm}$. A fully periodic system with $L=2.0032\times10^{-2}\text{cm}$ was initialized. Massively-parallel simulations on a $1024^3$ finite-volume grid were conducted for both deterministic Navier-Stokes and FHD on high-performance computing platforms (see Appendix \ref{HPC} for details). The finite-volume grid spacing $\Delta x = 1.956\times10^{-4}\text{mm}$ corresponds to $N\approx1.8\times10^{5}$ molecules of nitrogen per finite-volume cell. The time step of the simulation was fixed at $\Delta t=1.25\times10^{-11}\text{s}$ in both deterministic Navier-Stokes and FHD simulations. The thermodynamic and transport properties of the gas were modeled with a hard-sphere approximation based on the prescription by \citet{Giov1}. A turbulent solenoidal forcing corresponding to $\sigma = 6\times10^9\text{cm}/\text{s}^2$ and $T_L=1.5\times10^{-4}\text{s}$ was applied at the start to both deterministic Navier-Stokes and FHD simulations. In each case, the simulations were first run for about $1.125\times10^6$ time steps until they reached a statistical steady state. Thereafter, the simluations were run for at least longer than $8\tau_{\lambda}$ where $\tau_{\lambda}$ is the eddy turnover time during which the statistics were collected.

\begin{figure}%
    \centerline{\includegraphics[width=\columnwidth]{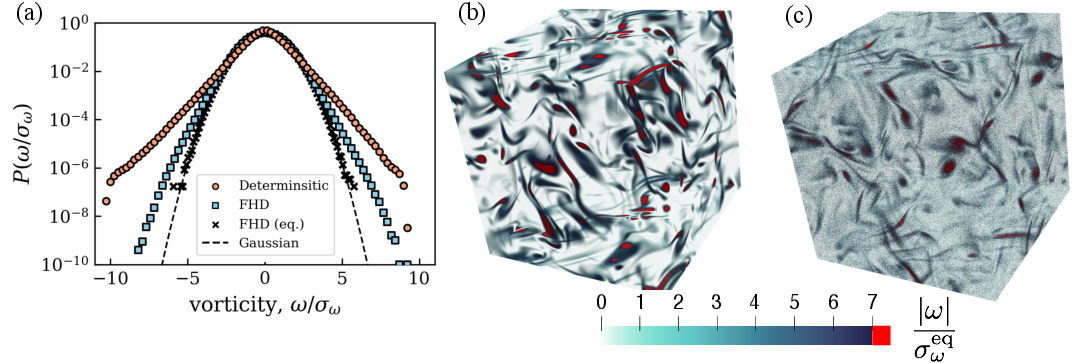}}
    \caption{(a) Probability distribution function (PDF) of local vorticity $\omega$ normalized by their ensemble standard deviation $\sigma_{\omega}$ averaged over at least $8\tau_{\lambda}$, where $\tau_{\lambda}$ is the eddy turnover time for deterministic and fluctuating hydrodynamics (FHD) simulations. The PDF from an FHD simulation at thermodynamic equilibrium without turbulent forcing, FHD (eq.), is also plotted. $3$D visualization of local vorticity magnitude $|\omega|$ in a (b) deterministic and an (c) FHD simulation. Here, $|\omega|$ is normalized by the standard deviation of vorticity fluctuations at thermodynamic equilibrium $\sigma_{\omega}^{\text{eq}}\approx5\times10^6\text{s}^{-1}$; the standard deviation of vorticity fluctuations $\sigma_{\omega}\approx7.3\times10^6\text{s}^{-1}$ and $\sigma_{\omega}\approx6.3\times10^6\text{s}^{-1}$ for deterministic and FHD simulations respectively.}    
\label{fig1}
\end{figure}

\section{Results}
\subsection{Dissipation-range turbulence with molecular fluctuations}
We first probe dissipation-range intermittency by analyzing the probability density function (PDF) of local vorticity obtained from direct numerical simulations (see Sec.~I of the Supplemental Material for details on the numerical computation of local vorticity) averaged over at least $8\tau_\lambda$, where $\tau_\lambda$ is the eddy turnover time. Intermittency in turbulent flows results in extreme bursts of local vorticity that are spatially interspersed within regions of relatively quiescent flow; as a result, the statistics of vorticity become highly non-Gaussian \citep{Frisch1995}. This is confirmed in figure \ref{fig1}(\emph{a}) that shows non-Gaussian tails in the PDF of the vector components of local vorticity $\omega$ normalized by the ensemble standard deviation $\sigma_{\omega}$. Remarkably, when molecular fluctuations are included (labeled FHD), a more Gaussian-like PDF is obtained that indicates the homogenizing effect of molecular fluctuations at dissipation scales that are about $3$ times larger than the molecular mean free path. In FHD simulations at thermodynamic equilibrium in the absence of external turbulent forcing, the PDF is completely Gaussian. For this case, the ensemble standard deviation of local vorticity $\sigma_{\omega}^{\text{eq}}$ matches well with theoretical predictions of equilibrium thermodynamics \citep{landau_statistical_1980}, to within less than $1\%$. The homogenizing effect of molecular fluctuations is readily observed in the visualization of local vorticity magnitude $|\omega|$ normalized by $\sigma_{\omega}^{\text{eq}}$ in figures \ref{fig1}(\emph{b}) and \ref{fig1}(\emph{c}). Whereas in deterministic simulations, regions of high vorticity are highly localized around large regions of quiescence, FHD simulations exhibit a more diffuse distribution of vorticity. Here, localized regions of high vorticity are overlaid on homogeneously distributed fluctuating velocity (and vorticity) as a result of thermal equipartition from molecular fluctuations. In FHD simulations at thermodynamic equilibrium, the local vorticity is a completely Gaussian random field ( (see Fig.~S1 of the Supplemental Material)).
\begin{figure}%
    \centerline{\includegraphics[width=\columnwidth]{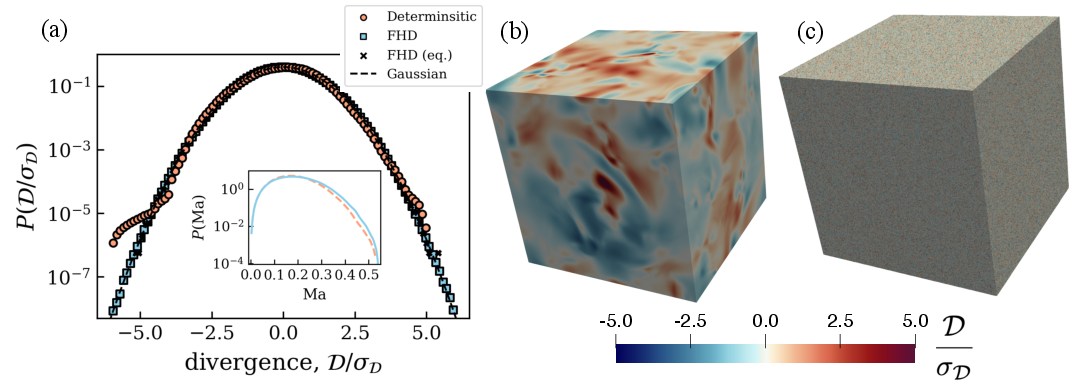}}
    \caption{(a) PDF of local divergence $\mathcal{D}$ normalized by its ensemble standard deviation $\sigma_{\mathcal{D}}$ for deterministic and fluctuating hydrodynamics (FHD) simulations. The inset in (a) shows the PDF of local Mach number $\Ma$ in FHD (orange) and deterministic (blue) simulations. $3$D visualization of local divergence in a (b) deterministic and an (c) FHD simulation. Here, $\mathcal{D}$ is normalized by the standard deviation of divergence fluctuations that are $\sigma_{\mathcal{D}}\approx3.1\times10^5\text{s}^{-1}$ and $\sigma_{\mathcal{D}}\approx8.7\times10^6\text{s}^{-1}$ for deterministic and FHD simulations respectively.}
\label{fig2}
\end{figure}

Compressible turbulence exhibits strong hydrodynamic shocks \citep{federrath2021sonic}; however, even weakly-compressible subsonic compressible turbulent flows can exhibit large density gradients without shocks \citep{benzi2008intermittency}. Here we restrict ourselves to nonlinear subsonic flows without any strong shock effects \citep{Sagaut2018}, but where the local Mach numbers can go as high as $0.5$ (see inset of figure \ref{fig2}(\emph{a})) such that compressibility effects are not negligible, and we observe regions of large density variations (see Fig.~S2 of the Supplemental Material for $3$D visualizations of local density fields). The dilatational behavior of turbulence is analyzed by the PDF of local divergence $\mathcal{D}=\bnabla\bcdot\uFluid$ normalized by the ensemble standard deviation $\sigma_{\mathcal{D}}$ in figure \ref{fig2}(\emph{a})  (see Sec.~I of the Supplemental Material for details on the numerical computation of local divergence). The PDF is nearly Gaussian for FHD simulations and is co-incident with the fully Gaussian PDF for FHD simulations without turbulent forcing. Deterministic simulations exhibit modest non-Gaussian tails for both positive and negative divergence. Furthermore, the instantaneous PDFs exhibit significant temporal variability in deterministic simulations whereas the variability is very small for FHD simulations (see Fig.~S3 of the Supplemental Material for PDFs of local divergence). On average, however, divergence in deterministic simulations is negatively skewed with skewness $\mathcal{S}\approx-0.12\pm0.19$, whereas $\mathcal{S}\approx0$ for FHD simulations. More spatial volume is associated with expansion than compression in deterministic simulations \citep{Sagaut2018}, whereas FHD simulations exhibit nearly equal volumes of expansion and compression. 

The strength of dilatation is much stronger in FHD simulations ($\sigma_{\mathcal{D}}\approx8.7\times10^6\text{s}^{-1}$) compared to deterministic simulations ($\sigma_{\mathcal{D}}\approx3.1\times10^5\text{s}^{-1}$). Molecular fluctuations in FHD simluations excite both vortical and dilatational modes of fluid motion via equipartition, whereas dilatational modes are indirectly excited through nonlinear coupling with the fluid vorticity in deterministic simulations \citep{Sagaut2018}, which is a much weaker effect for pure solenoidally-forced turbulent flows considered here. In FHD simulations with no turbulent forcing $\sigma_{\mathcal{D}}^{\text{eq}}\approx8.6\times10^6\text{s}^{-1}$, which nearly equal to its value in FHD simulations with turbulent forcing, thus demonstrating that molecular fluctuations completely dominate the dilatational dynamics. The differences are apparent in figures \ref{fig2}(\emph{b}) and \ref{fig2}(\emph{c}) that visualize local $\mathcal{D}/\sigma_{\mathcal{D}}$ fields. While deterministic simulations exhibit extended regions of both positive and negative divergence separated by contact discontinuities, the local divergence field is spatially nearly Gaussian in FHD simulations.

Here we remark that in order to derive various hydrodynamic quantities, such as vorticity and divergence discussed above, we computed the numerical derivatives of the velocity field on the finite-volume grid. We emphasize that the discrete numerical operators that are used to derive these quantities are the same operators that were used to evaluate derivatives in the numerical solution algorithm for the FHD equations, thus making them consistent with the underlying numerical algorithm.
As with the numerical solution of the FHD equations, the derived hydrodynamic quantities also depend on the mesh resolution; however, this resolution dependence is physically correct since the variance of thermal fluctuations depends on the scale of measurement.

\subsection{Turbulence and thermal dissipation: separation of scales}
\label{sec:turbchar}
\begin{figure}
    \centerline{\includegraphics[width=0.7\columnwidth]{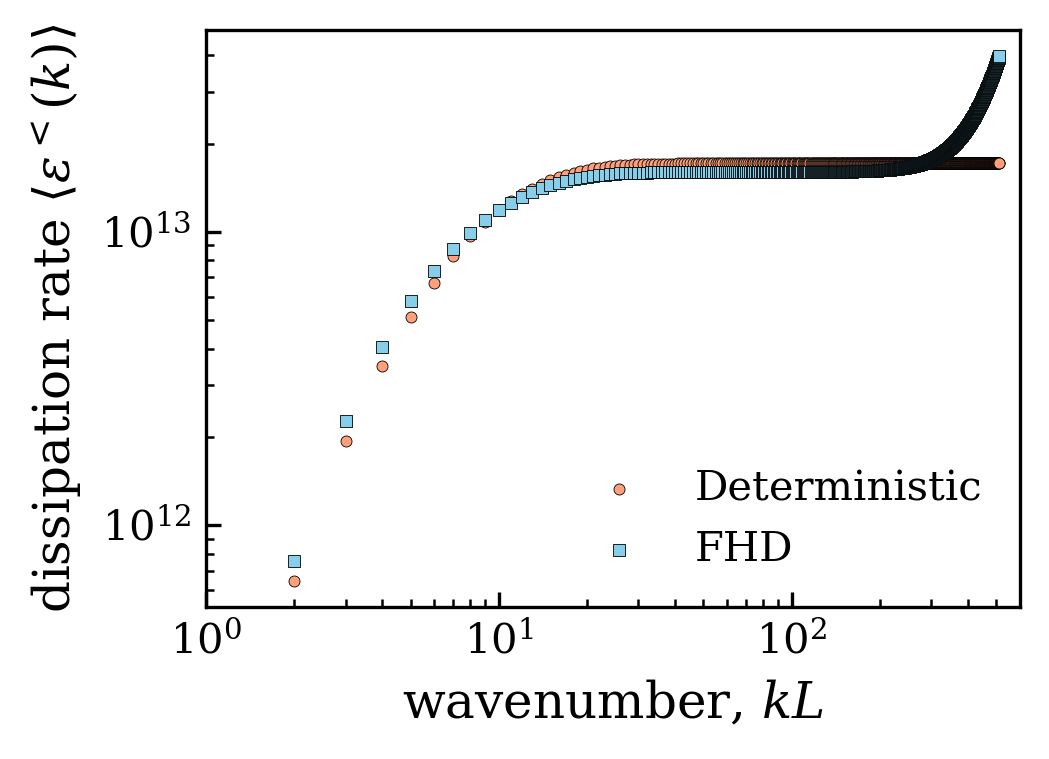}}
    \caption{Mean low-pass filtered dissipation rate $\langle \epsilon ^{<}(k)\rangle$ as a function of the wavenumber $k$ computed from the mean mean low-pass filtered enstrophy in Equation \ref{eq:enstrophy} for deterministic Navier-Stokes and FHD simulations of compressible turbulence.}
\label{fig:filter}
\end{figure}

In order to provide an objective comparison between deterministic Navier-Stokes and FHD simulations, we compute various microscale and dissipation (Kolmogorov)-scale turbulence quantities from the simulations. Unlike deterministic Navier-Stokes equations, the computation of velocity gradients in FHD is highly scale-dependent, and as such they do not represent an objective physical quantity. Therefore, any derived microscale and dissipation scale turbulence quantities from local velocity gradients will depend on the low-pass filter cutoff for the hydrodynamic and thermodynamic fields. In order to define an objective and meaningful turbulent energy dissipation rate, we compute the mean low-pass filtered enstrophy $\langle \Omega^{<}(k) \rangle$ from the kinetic energy spectrum $\langle E(k)\rangle =\frac{1}{2}\langle \hat{\uFluid}(k) \bcdot \hat{\uFluid}(k)^{*}\rangle$ as:
\begin{equation}
    \langle \Omega^{<}(k) \rangle = \int_{0}^{k} q^{2}\langle E(q) \rangle dq,
\label{eq:enstrophy}
\end{equation}
where $\hat{\uFluid}(k)$ is the total velocity in the Fourier space. Subsequently, a mean low-pass filtered dissipation rate is computed as $\langle \epsilon ^{<}(k)\rangle = \frac{2\langle\eta\rangle}{\langle\rho\rangle} \Omega^{<}(k)$.
Figure \ref{fig:filter} shows $\langle \epsilon ^{<}(k)\rangle$ as a function of the filtering wavenumber for deterministic Navier-Stokes and FHD simulations. At large cutoff wavenumbers, $\langle \epsilon ^{<}(k)\rangle$ plateaus to a constant value owing to very small velocities at small scales, whereas in FHD simulations $\langle \epsilon ^{<}(k)\rangle$ plateaus to a nearly similar constant value before rapidly increasing at even higher wavenumbers. This increase is attributed to dissipation primarily occurring from molecular fluctuations at small scales, which is a distinct effect than turbulent eddy fluctuations \citep{eyink2022high}. As such, the plateau value of $\langle \epsilon ^{<}(k)\rangle$, hereby denoted by $\langle \epsilon ^{<}\rangle$, provides a physically meaningful and objective definition of turbulent energy dissipation rate in deterministic Navier-Stokes and FHD simulations. Furthermore, the current experimental techniques for turbulence measure coarse-grained fluid velocities and dissipation rates at scales much larger than the Kolmogorov scale, and as such, are consistent with the low-pass filtered definition of these quantities. Future experiments that can measure sub-Kolmogorov-scale velocities can potentially disentangle dissipation due to molecular fluctuations from turbulence dissipation  \citep{bandak2022dissipation}.
\begin{table}
   \begin{center}
\def~{\hphantom{0}}
  \begin{tabular}{p{0.1\textwidth}p{0.1\textwidth}p{0.1\textwidth}p{0.1\textwidth}p{0.1\textwidth}p{0.1\textwidth}p{0.1\textwidth}}
\textrm{case}&
\textrm{$\Ma_t$}&
\textrm{$\Rey_{\lambda}$}&
\textrm{$l_{\lambda}\times10^{-3}$ (cm)}&
\textrm{$\tau_{\lambda}\times10^{-7}$ (s)}&
\textrm{$l_{\eta}\times10^{-4}$ (cm)}&
\textrm{$\tau_{\eta}\times10^{-7}$ (s)}\\
D-NS & $0.20$ & $34.9$ & $1.53$ & $4.13$ & $1.28$ & $1.01$  \\
FHD  & $0.21$ & $40.1$ & $1.67$ & $4.27$ & $1.26$ & $0.97$
\end{tabular}
  \caption{Mean turbulence statistics obtained from the simulations. D-NS and FHD denote deterministic Navier-Stokes and fluctuating hydrodynamics respectively. Here, $\Ma_t$ is the turbulent Mach number, $\Rey_{\lambda}$ is the microscale Reynolds number, $l_{\lambda}$ is the Taylor microscale length, $\tau_{\lambda}$ is the eddy turnover time, $l_{\eta}$ is the Kolmogorov length corresponding to the total dissipation rate, and $\tau_{\eta}$ is the Kolmogorov time scale.}
  \label{table1}
   \end{center}
\end{table}

Using the prescription for low-pass filtered turbulent energy dissipation rate discussed above, we derived various microscale and dissipation-scale quantities from the simulations.  In particular, we computed the following microscale quantities: (1) turbulent Mach number $\Ma_t=u'/\langle c \rangle $, where $c$ is the local speed of sound and $u'$ is the r.m.s. velocity that is computed from the kinetic energy spectrum as:
\begin{equation}
    u'^2 = \frac{2}{3}\int_{0}^{\infty}\langle E(k) \rangle dk;
\end{equation}
(2) microscale Reynolds number $\Rey_{\lambda} = \langle \rho \rangle u' l_{\lambda} /\langle \eta \rangle$ corresponding to the Taylor microscale length \citep{PopeTurbulentFlows2001}:
\begin{equation}
    l_{\lambda} = \sqrt{\frac{2u^{'2}}{\left\langle \left(\frac{\partial u_1}{\partial x_1}\right)^2\right\rangle}},
\end{equation}
where $\left\langle \left(\frac{\partial u_1}{\partial x_1}\right)^2\right\rangle = \frac{2}{9} \langle \Omega^{<}\rangle$ assuming isotropy of flow. Per the discussion above, we use the plateau value of $\langle \Omega^{<}\rangle$ to estimate the velocity gradients. A microscale eddy turnover time is also computed as $\tau_{\lambda}=l_{\lambda}/u'$. To compute dissipation-scale quantities, we use the plateau value of mean low-pass filtered dissipation rate $\langle \epsilon ^{<}\rangle$ as described above. The dissipation (Kolmogorov) length scale is calculated as $l_{\eta} = \left(\langle\eta\rangle^3 / \langle\rho\rangle^3 \langle\epsilon^{<}\rangle \right)^{1/4}$, and the corresponding Kolmogorov (small eddy turnover) time scale is calculated as $\tau_{\eta} = \left(\langle\eta\rangle / \langle\rho\rangle\langle\epsilon^{<}\rangle \right)^{1/2}$.

Table \ref{table1} lists the microscale and dissipation-scale turbulence statistics for deterministic Navier-Stokes and FHD simluations. By using a low-pass filter for velocity gradients and dissipation rates in the Fourier space as described above, we obtain a meaningful comparison between the two simulations. We note that in the case of deterministic simulation, the turbulent Mach number $\Rey_{\lambda}$ computed above matches reasonably well with its value of $\Rey_{\lambda}=41.8$ computed directly from the velocities in the real space on the finite-volume grid. The small discrepancy between the two values is possibly attributed to complex enstrophy budgeting among the nonlinearly coupled dilatational and solenoidal components of the turbulence velocity.

\subsection{Thermal energy crossover scale in the energy spectrum}
\label{sec:crossover}
\begin{figure}
    \centerline{\includegraphics[width=\columnwidth]{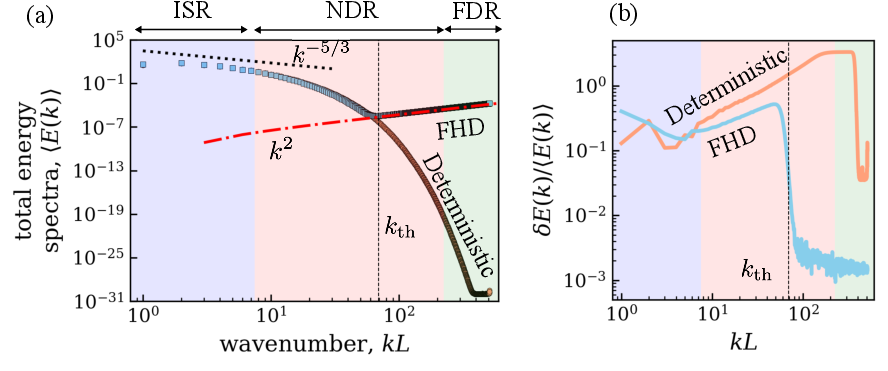}}
    \caption{(a) Comparison of the total kinetic energy spectrum $\langle E(k) \rangle$ in FHD vs. deterministic simulations. Three approximate ranges of length scales are highlighted: inertial sub-range (ISR in blue), near-dissipation range (NDR in pink) and far-dissipation range (FDR in green). In FHD simulations the thermal spectrum $E_{\text{th}}(k) =  \frac{3k_B \langle T \rangle}{2\langle \rho \rangle} 4\pi k^{2}$ (red dashed-dot line) dominates for wavenumbers larger than the thermal crossover scale $k_{\text{th}}$, where $k_B$ is the Boltzmann constant. (b) Standard deviation in total kinetic energy spectrum $\delta E(k) = \langle \left(E(k) - \langle E(k)\rangle \right)^{2} \rangle^{1/2}$ normalized by $\langle E(k) \rangle$.}
\label{fig3}
\end{figure}
We now discuss the length scales at which molecular fluctuations have an appreciable influence on compressible turbulence beyond the dissipation scale. The total energy spectra $E(k)=\frac{1}{2}\langle \hat{\uFluid}(k) \bcdot \hat{\uFluid}(k)^{*}\rangle$ of a turbulent flow can be approximately divided into the following three regimes (see figure \ref{fig3}(\emph{a})). (1) The far-dissipation range (FDR) represents the smallest length scales, specifically wavenumbers larger than the Kolmogorov wavenumber $k_{\eta} = \nu^{-3/4}\langle\epsilon\rangle^{1/4}$, where $\nu$ is the kinematic viscosity and $\langle\epsilon\rangle$ is the total mean dissipation rate. This regime is dominated by viscous dissipation and strong intermittency \citep{Kraichnan1967}, and molecular fluctuations strongly dominate turbulence at these length scales, as shown above. (2) The inertial sub-range (ISR) represents length scales where energy cascades from larger eddies to smaller eddies in a scale-invariant manner, and energy spectra has the form $E(k) \propto \langle\epsilon\rangle^{2/3}k^{-5/3}$ \citep{Frisch1995}. (3) The near-dissipation range (NDR) \citep{frisch1991prediction,Buaria2020PRF} that extends approximately from $k_{\eta}/30$ to $k_{\eta}$ represents the transition between ISR and FDR where the viscous effects start to become important and intermittency starts growing rapidly \citep{chevillard2005rapid}. Here, the turbulent spectra drops exponentially as $E(k) = u_{\eta}^{2} l_{\eta} \text{exp}\left(-\beta k l_{\eta}\right)$, where $u_{\eta} = \left(\langle \epsilon \rangle \nu\right)^{1/4}$ is the Kolmogorov velocity scale, $l_{\eta} = \left(\nu^{3}/\langle\epsilon\rangle\right)^{1/4}$ is the Kolmogorov length, and $\beta$ is the rate of exponential decay of the spectrum that typically ranges from $3-7$ (we have fixed $\beta=5$ in our analysis) \citep{Khurshid2018PRF}.

Molecular fluctuations introduce another length scale in the turbulence spectrum \citep{bandak2021}. From equilibrium thermodynamics, the contribution of molecular fluctuations to the energy spectrum (assuming no net flow, i.e., $\langle \uFluid\rangle = 0$) is:
\begin{equation}
    E_{\text{th}}(k) = \frac{3k_B \langle T \rangle}{2\langle \rho \rangle} 4\pi k^{2},
\end{equation}
which is `equipartitioned' white noise with a variance of $\frac{3k_B \langle T \rangle}{2\langle \rho\rangle}$ at all scales. The wavenumber $k_{\text{th}}$ at which molecular fluctuations are approximately equal in magnitude to the turbulent spectrum is \citep{bandak2021}:
\begin{equation}
    u_{\eta}^{2} l_{\eta} \text{exp}\left(-\beta k_{\text{th}} l_{\eta}\right) \approx \frac{k_B \langle T \rangle}{\langle \rho \rangle} k_{\text{th}}^{2}.
\label{eq:kcross}
\end{equation}

\begin{figure}
    \centerline{\includegraphics[width=\columnwidth]{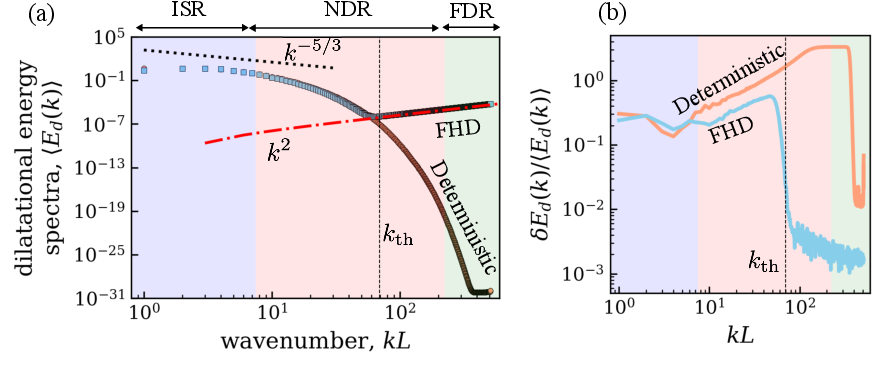}}
    \caption{(a) Comparison of dilatational kinetic energy $\langle  E_d(k) \rangle $ in FHD vs. deterministic simulations. The FHD simulations transitions over to the thermal energy spectrum is $E_{d,\text{th}}(k)=(1/3)E_{\text{th}}(k)$ (red dashed-dot line) at $k_{\text{th}}$. (b) Standard deviation in the dilatational kinetic energy spectrum $\delta E_d(k) = \langle \left(E_d(k) - \langle E_d(k)\rangle \right)^{2} \rangle^{1/2}$ normalized by $\langle E_d(k) \rangle$.}
\label{fig4}
\end{figure}

Indeed in figure \ref{fig3}(\emph{a}) we observe that for FHD simulations, the total energy spectrum crosses over from an exponential decay in the NDR to being dominated by the thermal spectrum $E_{\text{th}}(k)$ at high wavenumbers. The agreement with $E_{\text{th}}(k)$ is remarkable without any fitting parameters. The thermal crossover wavenumber $k_{\text{th}}$ is approximately three times smaller than the Kolmogorov wavenumber $k_{\eta}$, and its predicted value from (\ref{eq:kcross}) (shown by dashed vertical black line) matches well with the observed crossover to $E_{\text{th}}(k)$ (shown by dash-dot red line). While the ratio $k_{\text{th}}/k_{\eta}$ depends on turbulence conditions, such as density, viscosity, temperature and mean dissipation rate \citep{bandak2022dissipation,Bell2022thermal}, the relationship between $k_{\text{th}}$ and $k_{\eta}$ is fairly robust and varies only very marginally across a wide range of turbulence conditions \citep{Bell2022thermal}.

The crossover into the thermal regime is also observed for dilatational part of the energy spectrum $E_{d}(k)=\frac{1}{2}\langle \hat{\uFluid}_{d}(k) \bcdot \hat{\uFluid}_{d}(k)^{*}\rangle$, as shown in figure \ref{fig4}(\emph{a}), where $\hat{\uFluid}_{d}$ is the dilatational (curl-free) part of the total velocity $\hat{\uFluid}$. At low wavenumbers, the total kinetic energy is dominated by solenoidal modes since the external turbulence forcing is solenoidal (see Fig.~S4 of the Supplemental Material for $\langle E_d(k)/E(k) \rangle$). However, following a rapid decay in the NDR, $E_{d}(k)$ crosses over to $E_{d,\text{th}}(k)=(1/3)E_{\text{th}}(k)$ at the wavenumber $k_{\text{th}}$ in FHD simulations. The factor $1/3$ appears because one-third of the thermal energy of molecular fluctuations is `equipartitioned' into the dilatational part and two-thirds into the solenoidal part of the total kinetic energy.

The picture that emerges from these observations is that the impact of molecular fluctuations on turbulence is not limited to dissipation scales in the FDR, but appears at larger thermal crossover scales in the NDR. While the simulations in this study have been conducted at low Reynolds numbers due to computational constraints, we can estimate the scales at which molecular fluctuations will be significant in several practical scenarios. For example, following Refs.~\citep{Garratt1994,bandak2021}, in atmospheric boundary layer assuming to be composed entirely of nitrogen at $T=300\text{K}$, the energy dissipation rate is $\epsilon=400\text{cm}^{2}/\text{s}^{3}$, kinematic viscosity of nitrogen is $\nu=0.16\text{cm}^{2}/\text{s}$, density $\rho=1.1\times 10^{-3}\text{g}/\text{cm}^3$. The mean free path $l_{\text{mfp}}\approx70\text{nm}$, while the Kolmogorov length scale $l_{\eta}=0.57\text{mm}$. From Eq. 3.2, the thermal crossover length scale at which molecular fluctuations will dominate is $l_{\text{th}}\approx1.3\text{mm}$, which is over four orders of magnitude larger than the mean free path.

\subsection{Molecular fluctuations impact turbulence statistics across the near-dissipation range}
It is apparent that mean turbulence properties are significantly modified in the NDR at all length scales smaller than $1/k_{\text{th}}$. However, it is well-known that intermittency in turbulence starts building up in the ISR and rapidly increases in the NDR where viscous effects start to intensify \citep{frisch1991prediction,chevillard2005rapid}. Therefore, even though molecular fluctuations do not affect the ensemble averaged turbulence properties such as the energy spectrum $\langle E(k) \rangle$ for $k<k_{\text{th}}$, we can expect them to modify the statistical properties of turbulence.

Indeed, a remarkable picture emerges where the large temporal statistical variability of turbulence in the NDR is significantly \emph{reduced} due to molecular fluctuations. Figures \ref{fig3}(\emph{b}) and \ref{fig4}(\emph{b}) respectively show the standard deviation of the total energy $\delta E(k)$ and dilatational energy spectra $\delta E_d(k)$ normalized by the mean value averaged over at least $8\tau_{\lambda}$. The growth of $\delta E(k)$ and $\delta E_d(k)$ is much slower in FHD than deterministic simulations for $k<k_{\text{th}}$, thus implying increased statistical stability of the dynamical turbulent system with molecular fluctuations. For $k>k_{\text{th}}$, the statistical variability plummets by two orders of magnitude in FHD simulations whereas it keep increasing with $k$ for deterministic simulations up to the beginning of the FDR. The eventual drop-off in $\delta E(k)$ and $\delta E_d(k)$ at very high $k$ results from limitations in numerical precision.

Next, we quantify scale-dependent spatial intermittency of turbulence through high-pass filtered skewness $\mathcal{S}^{>}(k)$ and kurtosis (flatness) $\mathcal{K}^{>}(k)$ of the velocity gradient $\partial_{x} \uFluid^{>}$ that are computed as:
\begin{equation}
    \mathcal{S}^{>}(k_i) = \frac{\overline{\left(\partial_{x} \uFluid^{>}\right)^{3}}}{\left[\overline{\left(\partial_{x} \uFluid^{>}\right)^{2}}\right]^{3/2}}; \quad \quad \mathcal{K}^{>}(k_i) = \frac{\overline{\left(\partial_{x} \uFluid^{>}\right)^{4}}}{\left[\overline{\left(\partial_{x} \uFluid^{>}\right)^{2}}\right]^{2}},
\end{equation}
where
\begin{equation}
    \overline{\left(\partial_{x} \uFluid^{>}\right)^{n}} = \frac{1}{V}\int \text{d}\mathbf{r} \left(\partial_{x} \uFluid^{>}(\mathbf{r})\right)^{n},
\end{equation}
and $\uFluid^{>}$ is the high-pass filtered velocity. Numerically, $\uFluid^{>}$ is obtained by first computing the discrete Fourier transform of the velocity field over the finite volume grid and zeroing out the Fourier modes for wavenumbers smaller than $k$, followed by a discrete inverse Fourier transform to obtain the high-pass filtered velocity on the same finite volume grid. Once $\uFluid^{>}$ is obtained, $\mathcal{S}^{>}(k)$ and $\mathcal{K}^{>}(k)$ are calculated by numerically computing the derivative $\partial_{x} \uFluid^{>}$ using the same gradient operators as employed in the numerical simluation of the FHD equations. 

In an intermittent dynamical system, $\mathcal{K}^{>}(k)$ is expected to grow unbounded with $k$ in the NDR and into FDR as regions of intense turbulent activity become increasingly localized in smaller fractions of the system volume \citep{Frisch1995}. A negative skewness for a turbulent system implies energy cascade from large to small scales \citep{Frisch1995}, and its magnitude ranges from approximately $\mathcal{S}\approx-0.5$ to $\mathcal{S}\approx-0.3$. In a fully Gaussian distribution, $\mathcal{S}=0$ and $\mathcal{K}=3$.

\begin{figure}
\centerline{\includegraphics[width=\columnwidth]{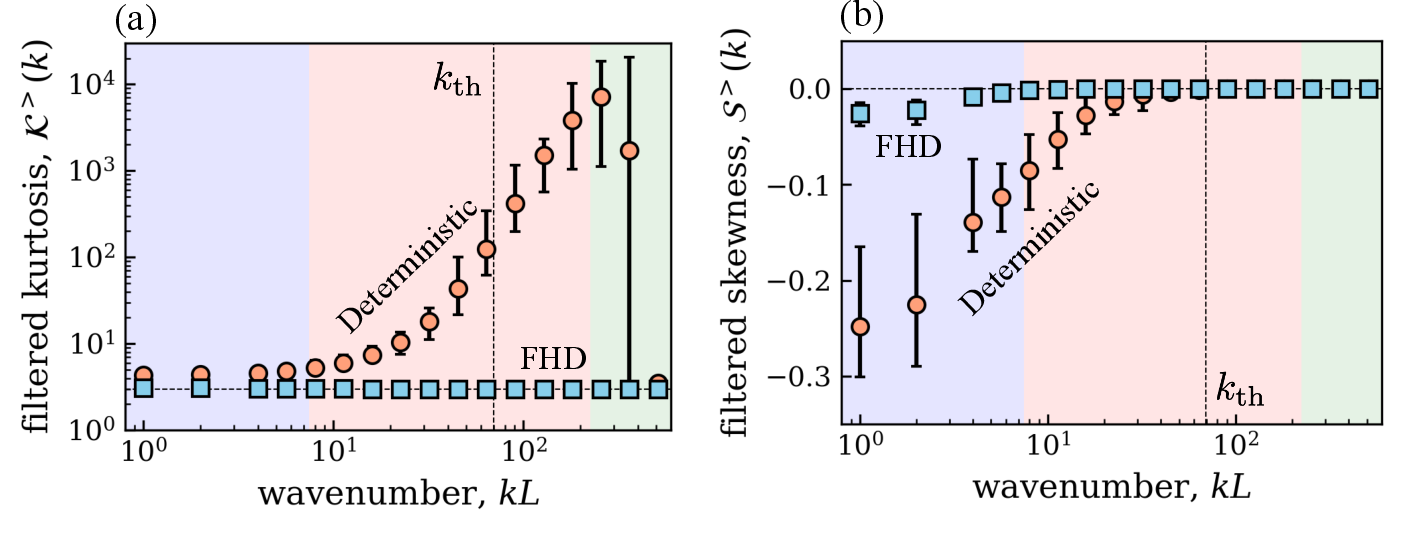}}
\caption{(a) Filtered kurtosis (flatness) $\mathcal{K}^{>}(k)$ and (b) filtered skewness $\mathcal{S}^{>}(k)$ of the velocity gradient $\partial_{x} \uFluid^{>}$, where $\uFluid^{>}$ is high-pass filtered velocity obtained by zeroing out all the Fourier modes for wavenumbers lesser than $k$ in the velocity field. The horizontal dashed line corresponds to the kurtosis and skewness of a Gaussian random field with $\mathcal{K}^{>}=3$ and $\mathcal{S}^{>}=0$ for all wavenumbers. The errors bars denote the ensemble standard deviation.}
\label{fig5}
\end{figure}

In the present simulations, rapidly increasing intermittency from its buildup in the ISR and propagation through the NDR and into FDR is observed in the deterministic case, as seen by the variation of $\mathcal{K}^{>}$ in figure \ref{fig5}(\emph{a}). In a remarkable contrast, $\mathcal{K}^{>}(k)\approx3$ at all wavenumbers in FHD simulations, thus demonstrating that the intermittent dynamics are completely inhibited not just in the FDR but well into the NDR. Furthermore, large variations in $\mathcal{K}^{>}(k)$ in deterministic simulations at high $k$, which are indicative of highly intermittent behavior, are not observed in FHD simulations. On the other hand, the skewness of velocity gradient $\mathcal{S}^{>}(k)$ in figure  \ref{fig5}(\emph{b}) saturates to its Gaussian value, as expected, for both FHD and deterministic simulations at high $k$. However at low $k$, deterministic simulations exhibit a negative skewness with large variability, whereas it is of a much smaller magnitude and variability in FHD simulations. We note that in a recent study on the role of molecular fluctuations in incompressible turbulence \citep{Bell2022thermal}, the skewness and kurtosis of the velocity gradient were reported to be unaffected by molecular fluctuations. Furthermore, through the analysis of structure functions in recent studies on stochastic shell modeling of incompressible turbulence \citep{bandak2022dissipation} and molecular gas dynamics simulations of compressible turbulence \citep{mcmullen2023thermal}, it was observed that while the far-dissipation range intermittency is replaced by
Gaussian fluctuations, the intermittency in the intermediate range persists. While our results are consistent with the studies in the far-dissipation range, our observations of drastically reduced intermittency in the near-dissipation range can potentially be attributed to low-$\Rey$ flows simulated here and/or compressibility effects.

\begin{figure}
\centerline{\includegraphics[width=\columnwidth]{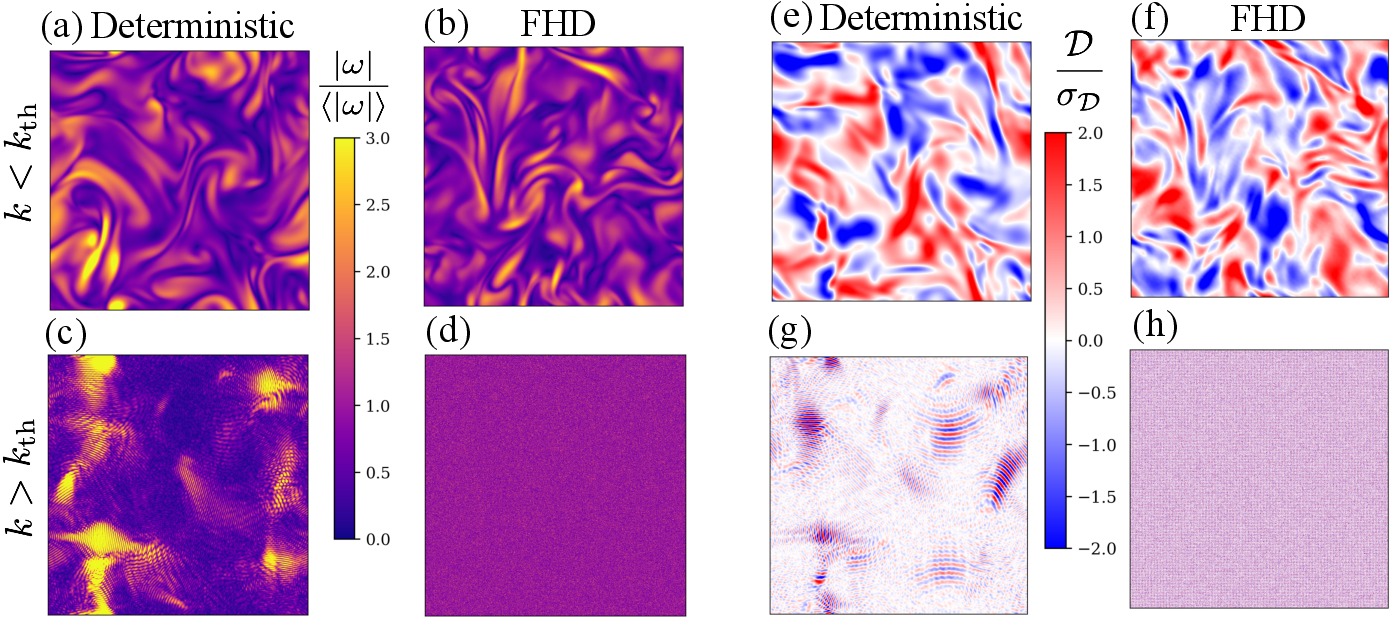}}
\caption{Cross-sectional visualization of the local vorticity magnitude $|\omega|$ (normalized by the ensemble mean $\langle|\omega|\rangle$) only for wavenumbers $k<k_{\text{th}}$ in a (a) deterministic and an (b) FHD simulation. (c) and (d) Same as (a) and (b) respectively, but only for wavenumbers $k>k_{\text{th}}$.
Cross-sectional visualization of the local divergence $\mathcal{D}$ (normalized by the ensemble standard deviation $\sigma_{\mathcal{D}}$) only for wavenumbers $k<k_{\text{th}}$ in a (e) deterministic and an (f) FHD simulation. (g) and (h) Same as (e) and (f) respectively, but only for wavenumbers $k>k_{\text{th}}$.}
\label{fig6}
\end{figure}

A visual analysis of the filtered invariants of velocity gradient (i.e., vorticity magnitude $|\omega|$ and divergence $\mathcal{D}$) highlights our observations. Figures \ref{fig6}(\emph{a}) and \ref{fig6}(\emph{b}) show $2$D slices of vorticity magnitude $|\omega|$, and figures \ref{fig6}(\emph{e}) and \ref{fig6}(\emph{f}) show $2$D slices of divergence $\mathcal{D}$ filtered for wavenumbers $k<k_{\text{th}}$. Similarly, figures \ref{fig6}(\emph{c}) and \ref{fig6}(\emph{d}), and figures \ref{fig6}(\emph{g}) and \ref{fig6}(\emph{h}), show the same data but filtered for wavenumbers $k>k_{\text{th}}$. While these fields `appear' similar at large wavelengths, $k<k_{\text{th}}$, in FHD and deterministic simulations, the visual differences are significant wavenumbers $k>k_{\text{th}}$. Here, FHD simulations exhibit a nearly homogeneous spatial distribution of vorticity and divergence with no signs of intermittency, whereas deterministic simulations exhibit classic signs of dissipation-range intermittency with localized bursts of high vorticity and divergence in a `sea' of quiescent fluid.

\section{Discussion}
Our simulations demonstrate that molecular fluctuations fundamentally modify compressible turbulence across the entire dissipation range, both in the energy spectrum and significantly reduced spatio-temporal intermittency. We propose that compressible fluctuating hydrodynamics (FHD) equations are a more appropriate mathematical model for compressible turbulence than the Navier-Stokes equations, especially for modeling dissipation-range physics. While FHD equations assume local thermodynamic equilibrium, they have successfully modeled compressible flows with large density gradients that compared well with molecular gas dynamics that make no such assumption \citep{srivastava2023staggered}. Importantly, even for weakly-compressible turbulent flows, the present results correspond well with recent molecular gas dynamics simulations of decaying turbulence \citep{mcmullen2022navier}.
However, the validity of FHD in strongly compressible turbulent flows with hydrodynamic shocks remains to be established and is a significant mathematical challenge.

In principle, our predictions can be tested in experiments; however, most current experiments lack spatial and temporal resolution, and sensitivity, to accurately probe dissipation-range turbulence \citep{bandak2022dissipation}. While some recent advances appear promising \citep{van2022dispersion}, the role of molecular fluctuations in turbulence can also be indirectly evidenced in physical processes \citep{bandak2022dissipation}. For example, molecular fluctuations have large observed macroscale effects in laminar diffusive mixing \citep{GiantFluctuations_Nature} and reacting flows \citep{Detonation_Fluctuations}; we can expect that molecular fluctuations will also impact the turbulent form of these processes. However, existing models of turbulent mixing \citep{sreenivasan2019turbulent} and combustion \citep{sreenivasan2004possible} do not account for them. Molecular fluctuations can also play an important role in transition to turbulence \citep{Betchov1961}, and recent efforts have explored the receptivity of compressible boundary layer to molecular fluctuations with design implications for high-speed aircraft \citep{fedorov2015prediction,luchini2017receptivity}.

Our results motivate new theoretical developments in turbulence closure models \citep{zhou2021turbulence} that correctly account for molecular fluctuations and its impact on intermittency. Correspondingly, latest developments in computational FHD to model thermal noise in multicomponent \citep{srivastava2023staggered} and reactive \citep{polimeno2024thermodynamic} flows will facilitate a new class of direct numerical simulations that can utilize exascale supercomputers to directly investigate the role of molecular fluctuations in a variety of large-scale turbulent flows.

\backsection[Supplementary data]{\label{SupMat}Supplementary material is available.}

\backsection[Funding]{This work was supported by the U.S. Department of Energy (DOE), Office of Science, Office of Advanced Scientific Computing Research, Applied Mathematics Program under contract No. DE-AC02-05CH11231. This research used resources of the Oak Ridge Leadership Computing Facility at the Oak Ridge National Laboratory, which is supported by the Office of Science of the U.S. Department of Energy under Contract No. DE-AC05-00OR22725. This research also used resources of the National Energy Research Scientific Computing Center (NERSC), DOE Office of Science User Facility supported by the Office of Science of the U.S. Department of Energy under Contract No. DE-AC02-05CH11231 using NERSC award ASCR-ERCAP0026881.}

\backsection[Declaration of interests]{The authors report no conflict of interest.}

\backsection[Data availability statement]{The code/data that support the findings of this study are openly available in
the github repository: \hyperlink{}{https://github.com/AMReX-FHD/FHDeX}.}

\backsection[Author ORCIDs]{I. Srivastava, https://orcid.org/0000-0003-4754-3232; A. Nonaka, https://orcid.org/0000-0003-1791-0265; W. Zhang, https://orcid.org/0000-0001-8092-1974; A. Garcia, https://orcid.org/0000-0003-3477-5982; J. Bell, https://orcid.org/0000-0002-5749-334X}

\appendix

\section{High-performance computing}\label{HPC}
The numerical method described here is implemented within the AMReX framework \cite{zhang2019a}, which uses an MPI paradigm for massively-parallel simulations along with GPU-based performance acceleration. The numerical method has been implemented in the fluctuating hydrodynamics software, FHDeX, and it is available online as an open-source code at \hyperlink{FHDeX}{https://github.com/AMReX-FHD/FHDeX}.

Most of the simulations were performed on the exascale supercomputing platform, Frontier, at the Oak Ridge National Laboratory. Each simulation run utilized either $256$ or $512$ compute nodes of Frontier; each compute node has $64$-core AMD `Optimized 3rd Gen EPYC' CPUs and $4$ AMD Instinct MI$250$X GPUs, where each GPU features $2$ Graphics Compute Dies (GCDs) for a total of 8 GCDs per compute node. All the simulations were run for approximately $1.5\times10^6$ to $2\times10^6$ time steps including the initial run to reach the steady state followed by simulation runs to extract turbulence statistics. In total, approximately $15,000$ GPU-hours were utilized to perform the simulations and analysis in this work, and $\mathcal{O}(10^{2})$ terabytes of raw data was generated.

\bibliographystyle{jfm}
\bibliography{TurbFHD}

@article{eyink2024onsager,
  title={Onsager's ‘ideal turbulence’theory},
  author={Eyink, Gregory},
  journal={J. Fluid Mech.},
  volume={988},
  pages={P1},
  year={2024},
  publisher={Cambridge University Press}
}

@article{mcmullen2023thermal,
  title={Thermal-fluctuation effects on small-scale statistics in turbulent gas flow},
  author={McMullen, RM and Torczynski, JR and Gallis, MA},
  journal={Physics of Fluids},
  volume={35},
  number={1},
  year={2023},
  publisher={AIP Publishing}
}

@article{morozov1984langevin,
  title={On the Langevin formalism for nonlinear and nonequilibrium hydrodynamic fluctuations},
  author={Morozov, VG},
  journal={Physica A: Statistical Mechanics and its Applications},
  volume={126},
  number={3},
  pages={443--460},
  year={1984},
  publisher={Elsevier}
}

@article{polimeno2024thermodynamic,
  title={Thermodynamic consistency and fluctuations in mesoscopic stochastic simulations of reactive gas mixtures},
  author={Polimeno, Matteo and Kim, Changho and Blanchette, Fran{\c{c}}ois and Srivastava, Ishan and Garcia, Alejandro L and Nonaka, Andy J and Bell, John B},
  journal={arXiv preprint arXiv:2412.07048},
  year={2024}
}

@article{fedorov2015prediction,
  title={Prediction and control of laminar-turbulent transition in high-speed boundary-layer flows},
  author={Fedorov, Alexander V},
  journal={Procedia Iutam},
  volume={14},
  pages={3--14},
  year={2015},
  publisher={Elsevier}
}

@article{sreenivasan2004possible,
  title={Possible effects of small-scale intermittency in turbulent reacting flows},
  author={Sreenivasan, KR},
  journal={Flow Turbul. Combust.},
  volume={72},
  pages={115--131},
  year={2004},
  publisher={Springer}
}

@article{sreenivasan2019turbulent,
  title={Turbulent mixing: A perspective},
  author={Sreenivasan, Katepalli R},
  journal={Proc. Natl. Acad. Sci. U.S.A},
  volume={116},
  number={37},
  pages={18175--18183},
  year={2019},
  publisher={National Acad Sciences}
}

@article{van2022dispersion,
  title={Dispersion of molecular patterns written in turbulent air},
  author={Van De Water, Willem and Dam, Nico and Calzavarini, Enrico},
  journal={Phys. Rev. Lett.},
  volume={129},
  number={25},
  pages={254501},
  year={2022},
  publisher={APS}
}

@article{zhou2021turbulence,
  title={Turbulence theories and statistical closure approaches},
  author={Zhou, Ye},
  journal={Phys. Rep.},
  volume={935},
  pages={1--117},
  year={2021},
  publisher={Elsevier}
}

@Inbook{Sagaut2018,
author="Sagaut, Pierre
and Cambon, Claude",
title="Compressible Homogeneous Isotropic Turbulence",
bookTitle="Homogeneous Turbulence Dynamics",
year="2018",
publisher="Springer International Publishing",
pages="621--687"
}

@article{bender2021simulation,
  title={Simulation and flow physics of a shocked and reshocked high-energy-density mixing layer},
  author={Bender, Jason D and Schilling, Oleg and Raman, Kumar S and Managan, Robert A and Olson, Britton J and Copeland, Sean R and Ellison, C Leland and Erskine, David J and Huntington, Channing M and Morgan, Brandon E and others},
  journal={J. Fluid Mech.},
  volume={915},
  pages={A84},
  year={2021},
  publisher={Cambridge University Press}
}

@book{landau_statistical_1980,
	address = {Amsterdam},
	edition = {3rd},
	title = {Statistical {Physics} {Part} {I}},
	publisher = {Elsevier},
	author = {Landau, L. D. and Lifshitz, E. M.},
	year = {1980},
}

@article{frisch1991prediction,
  title={A prediction of the multifractal model: the intermediate dissipation range},
  author={Frisch, U and Vergassola, M},
  journal={Europhys. Lett.},
  volume={14},
  number={5},
  pages={439},
  year={1991},
  publisher={IOP Publishing}
}

@article{neumann1949recent,
  title={Recent theories of turbulence},
  author={von Neumann, J.},
  journal={Collected Works (1949-1963)},
  volume={6},
  pages={437--472},
  year={1949},
  publisher={Pergamon Press}
}

@Article{zhang2019a,
author = {Zhang, W. and Almgren, A. and Beckner, V. and Bell, J. and Blaschke, J. and Chan, C. and Day, M. and Friesen, B. and Gott, K. and Graves, D. and Katz, M. and Myers, A. and Nguyen, T. and Nonaka, A. and Rosso, M. and Williams, S. and Zingale, M. }, 
title = {Amrex: A Framework for Block-Structured Adaptive Mesh Refinement}, 
journal = {J. Open Source Softw.}, 
volume = {4}, 
number = {37}, 
pages = {1370}, 
year = {2019}
}

@book{Giov1,
  title={Multicomponent Flow Modeling},
  author={Giovangigli, V.},
  isbn={9781461215806},
  series={Modeling and Simulation in Science, Engineering and Technology},
  year={2012},
  number = {N/A}, 
  publisher={Birkhauser Boston}
}

@article{tremblay1981fluctuations,
  title={Fluctuations about simple nonequilibrium steady states},
  author={Tremblay, A-MS and Arai, M and Siggia, ED},
  journal={Phys. Rev. A},
  volume={23},
  number={3},
  pages={1451},
  year={1981},
  publisher={APS}
}

@article{urzay2018supersonic,
  title={Supersonic combustion in air-breathing propulsion systems for hypersonic flight},
  author={Urzay, Javier},
  journal={Ann. Rev. Fluid Mech.},
  volume={50},
  number={1},
  pages={593--627},
  year={2018},
  publisher={Annual Reviews}
}

@article{chevillard2005rapid,
  title={On the rapid increase of intermittency in the near-dissipation range of fully developed turbulence},
  author={Chevillard, Laurent and Castaing, Bernard and L{\'e}v{\^e}que, Emmanuel},
  journal={Eur. Phys. J. B},
  volume={45},
  number={4},
  pages={561--567},
  year={2005},
  publisher={Springer}
}

@article{hamlington2012intermittency,
  title={Intermittency in premixed turbulent reacting flows},
  author={Hamlington, Peter E and Poludnenko, Alexei Y and Oran, Elaine S},
  journal={Phys. Fluids},
  volume={24},
  number={7},
  year={2012},
  publisher={AIP Publishing}
}

@article{donzis2020universality,
  title={Universality and scaling in homogeneous compressible turbulence},
  author={Donzis, Diego A and John, John Panickacheril},
  journal={Phys. Rev. Fluids},
  volume={5},
  number={8},
  pages={084609},
  year={2020},
  publisher={APS}
}

@article{wang2013cascade,
  title={Cascade of kinetic energy in three-dimensional compressible turbulence},
  author={Wang, Jianchun and Yang, Yantao and Shi, Yipeng and Xiao, Zuoli and He, XT and Chen, Shiyi},
  journal={Phys. Rev. Lett.},
  volume={110},
  number={21},
  pages={214505},
  year={2013},
  publisher={APS}
}

@article{wang2012scaling,
  title={Scaling and statistics in three-dimensional compressible turbulence},
  author={Wang, Jianchun and Shi, Yipeng and Wang, Lian-Ping and Xiao, Zuoli and He, XT and Chen, Shiyi},
  journal={Phys. Rev. Lett.},
  volume={108},
  number={21},
  pages={214505},
  year={2012},
  publisher={APS}
}

@article{federrath2021sonic,
  title={The sonic scale of interstellar turbulence},
  author={Federrath, Christoph and Klessen, Ralf S and Iapichino, Luigi and Beattie, James R},
  journal={Nat. Astron.},
  volume={5},
  number={4},
  pages={365--371},
  year={2021},
  publisher={Nature Publishing Group UK London}
}

@article{eyink2018cascades,
  title={Cascades and dissipative anomalies in compressible fluid turbulence},
  author={Eyink, Gregory L and Drivas, Theodore D},
  journal={Phys. Rev. X},
  volume={8},
  number={1},
  pages={011022},
  year={2018},
  publisher={APS}
}

@article{aluie2011compressible,
  title={Compressible turbulence: the cascade and its locality},
  author={Aluie, Hussein},
  journal={Phys. Rev. Lett.},
  volume={106},
  number={17},
  pages={174502},
  year={2011},
  publisher={APS}
}

@article{mac2004control,
  title={Control of star formation by supersonic turbulence},
  author={Mac Low, Mordecai-Mark and Klessen, Ralf S},
  journal={Rev. Mod. Phys.},
  volume={76},
  number={1},
  pages={125},
  year={2004},
  publisher={APS}
}

@article{benzi2008intermittency,
  title={Intermittency and Universality in Fully Developed Inviscid and Weakly Compressible Turbulent Flows},
  author={Benzi, Roberto and Biferale, Luca and Fisher, Robert T and Kadanoff, Leo P and Lamb, Donald Q and Toschi, Federico},
  journal={Phys. Rev. Lett.},
  volume={100},
  number={23},
  pages={234503},
  year={2008},
  publisher={APS}
}

@article{wang2017scaling,
  title={Scaling and intermittency in compressible isotropic turbulence},
  author={Wang, Jianchun and Gotoh, Toshiyuki and Watanabe, Takeshi},
  journal={Phys. Rev. Fluids},
  volume={2},
  number={5},
  pages={053401},
  year={2017},
  publisher={APS}
}

@article{yeung2015extreme,
  title={Extreme events in computational turbulence},
  author={Yeung, PK and Zhai, XM and Sreenivasan, Katepalli R},
  journal={Proc. Natl. Acad. Sci. U.S.A},
  volume={112},
  number={41},
  pages={12633--12638},
  year={2015},
  publisher={National Acad Sciences}
}

@article{frisch1981intermittency,
  title={Intermittency in nonlinear dynamics and singularities at complex times},
  author={Frisch, Uriel and Morf, Rudolf},
  journal={Phys. Rev. A},
  volume={23},
  number={5},
  pages={2673},
  year={1981},
  publisher={APS}
}

@article{paladin1987anomalous,
  title={Anomalous scaling laws in multifractal objects},
  author={Paladin, Giovanni and Vulpiani, Angelo},
  journal={Phys. Rep.},
  volume={156},
  number={4},
  pages={147--225},
  year={1987},
  publisher={Elsevier}
}

@article{alexakis2018cascades,
  title={Cascades and transitions in turbulent flows},
  author={Alexakis, Alexandros and Biferale, Luca},
  journal={Phys. Rep.},
  volume={767},
  pages={1--101},
  year={2018},
  publisher={Elsevier}
}

@article{eyink2006onsager,
  title={Onsager and the theory of hydrodynamic turbulence},
  author={Eyink, Gregory L and Sreenivasan, Katepalli R},
  journal={Rev. Mod. Phys.},
  volume={78},
  number={1},
  pages={87--135},
  year={2006},
  publisher={APS}
}

@article{srivastava2023staggered,
  title={Staggered scheme for the compressible fluctuating hydrodynamics of multispecies fluid mixtures},
  author={Srivastava, Ishan and Ladiges, Daniel R and Nonaka, Andy J and Garcia, Alejandro L and Bell, John B},
  journal={Phys. Rev. E},
  volume={107},
  number={1},
  pages={015305},
  year={2023},
  publisher={APS}
}

@article{eyink2022high,
  title={High Schmidt-number turbulent advection and giant concentration fluctuations},
  author={Eyink, Gregory and Jafari, Amir},
  journal={Phys. Rev. Res.},
  volume={4},
  number={2},
  pages={023246},
  year={2022},
  publisher={APS}
}

@article{luchini2017receptivity,
  title={Receptivity to thermal noise of the boundary layer over a swept wing},
  author={Luchini, Paolo},
  journal={AIAA J.},
  volume={55},
  number={1},
  pages={121--130},
  year={2017},
  publisher={American Institute of Aeronautics and Astronautics}
}

@article{zubarev1983statistical,
  title={Statistical mechanics of nonlinear hydrodynamic fluctuations},
  author={Zubarev, DN and Morozov, VG},
  journal={Physica A},
  volume={120},
  number={3},
  pages={411--467},
  year={1983},
  publisher={Elsevier}
}

@article{Kraichnan1967,
  title={Intermittency in the very small scales of turbulence},
  author={Kraichnan, Robert H},
  journal={Phys. Fluids},
  volume={10},
  number={9},
  pages={2080--2082},
  year={1967},
  publisher={American Institute of Physics}
}

@incollection{Betchov1961,
   author={Betchov, R.},
   title={Thermal agitation and turbulence},
   booktitle={Rarefied Gas Dynamics}, 
   note={Proceedings of the Second International Symposium on Rarefied Gas Dynamics, held at the 
University of California, Berkeley, CA, 1960}, 
   editor={Talbot, L.},  
   publisher={Academic Press}, 
   address={New York},
   year={1961}, 
   pages={307–321} 
}

@article{Garratt1994,
  title={The atmospheric boundary layer},
  author={Garratt, John Roy},
  journal={Earth-Science Reviews},
  volume={37},
  number={1-2},
  pages={89--134},
  year={1994},
  publisher={Elsevier}
}

@book{Frisch1995,
  title={Turbulence: the legacy of AN Kolmogorov},
  author={Frisch, Uriel},
  year={1995},
  publisher={Cambridge university press}
}

@article{Betchov1957,
  title={On the fine structure of turbulent flows},
  author={Betchov, R},
  journal={J. Fluid Mech.},
  volume={3},
  number={2},
  pages={205--216},
  year={1957},
  publisher={Cambridge University Press}
}

@article{bandak2022dissipation,
  title={Dissipation-range fluid turbulence and thermal noise},
  author={Bandak, Dmytro and Goldenfeld, Nigel and Mailybaev, Alexei A and Eyink, Gregory},
  journal={Phys. Rev. E},
  volume={105},
  number={6},
  pages={065113},
  year={2022},
  publisher={APS}
}

@article{bandak2021,
  title={Thermal noise competes with turbulent fluctuations below millimeter scales},
  author={Bandak, Dmytro and Eyink, Gregory L and Mailybaev, Alexei and Goldenfeld, Nigel},
  journal={arXiv preprint arXiv:2107.03184},
  year={2021}
}

@article{Bell2022thermal,
  title={Thermal fluctuations in the dissipation range of homogeneous isotropic turbulence},
  author={Bell, John B and Nonaka, Andrew and Garcia, Alejandro L and Eyink, Gregory},
  journal={J. Fluid Mech.},
  volume={939},
  pages={A12},
  year={2022},
  publisher={Cambridge University Press}
}

@article{mcmullen2022navier,
  title={Navier-Stokes equations do not describe the smallest scales of turbulence in gases},
  author={McMullen, Ryan M and Krygier, Michael C and Torczynski, John R and Gallis, Michael A},
  journal={Phys. Rev. Lett.},
  volume={128},
  number={11},
  pages={114501},
  year={2022},
  publisher={APS}
}

@book{ZarateBook2006,
  title={Hydrodynamic fluctuations in fluids and fluid mixtures},
  author={De Zarate, Jose M Ortiz and Sengers, Jan V},
  year={2006},
  publisher={Elsevier}
}

@article{DonevPRL2011,
  title={Diffusive transport by thermal velocity fluctuations},
  author={Donev, Aleksandar and Bell, John B and de La Fuente, Anton and Garcia, Alejandro L},
  journal={Phys. Rev. Lett.},
  volume={106},
  number={20},
  pages={204501},
  year={2011},
  publisher={APS}
}

@book{PopeTurbulentFlows2001,
  title={Turbulent flows},
  author={Pope, Stephen B},
  year={2001},
  number={},
  publisher={IOP Publishing}
}

@article{Khurshid2018PRF,
  title={Energy spectrum in the dissipation range},
  author={Khurshid, Sualeh and Donzis, Diego A and Sreenivasan, KR},
  journal={Phys. Rev. Fluids},
  volume={3},
  number={8},
  pages={082601},
  year={2018},
  publisher={APS}
}

@BOOK{Landau1959Fluid,
   author       = {L. D. Landau and E. M. Lifshitz},
   year         = 1959,
   title        = {Fluid Mechanics, Course of Theoretical Physics, Vol. 6},
   publisher    = {Pergamon Press}
}

@article{Donev2010Camcos,
  author = "A. Donev and E. Vanden-Eijnden and A. Garcia and J. Bell",
  title = "On the accuracy of finite-volume schemes for fluctuating hydrodynamics",
  journal = "Commun. Appl. Math. Comput. Sci.",
  volume = "5",
  pages = "149--197",
  year = "2010"
}

@article{Buaria2020PRF,
  title={Dissipation range of the energy spectrum in high Reynolds number turbulence},
  author={Buaria, Dhawal and Sreenivasan, Katepalli R},
  journal={Physical Review Fluids},
  volume={5},
  number={9},
  pages={092601},
  year={2020},
  publisher={APS}
}

@book{Gardiner1985book,
  title={Handbook of stochastic methods},
  author={Gardiner, Crispin W.},
  volume={3},
  year={1985},
  publisher={springer Berlin}
}

@article{Eswaran1988CompFluids,
  title={An examination of forcing in direct numerical simulations of turbulence},
  author={Eswaran, Vinayak and Pope, Stephen B},
  journal={Comput. Fluids},
  volume={16},
  number={3},
  pages={257--278},
  year={1988},
  publisher={Elsevier}
}

@ARTICLE{LLNS_Staggered,
  author = {F. Balboa Usabiaga and J. B. Bell and R. Delgado-Buscalioni and A. Donev and T. G. Fai and B. E. Griffith and C. S. Peskin},
  title = {{Staggered Schemes for Fluctuating Hydrodynamics}},
  journal = {Multiscale Model. Sim.},
  year = {2012},
  volume = {10},
  number = {4},
  pages = {1369-1408},
}

@ARTICLE{PhysRev.187.267,
  author = {Bixon, Mordechai and Zwanzig, Robert },
  title = {Boltzmann-Langevin Equation and Hydrodynamic Fluctuations},
  journal = {Phys. Rev.},
  year = {1969},
  volume = {187},
  pages = {267--272},
  number = {1},
  numpages = {5},
  publisher = {American Physical Society}
}

@ARTICLE{DiscreteLLNS_Espanol,
  author = {Espa{\~n}ol, P. and Anero, J.G. and Z{\'u}{\~n}iga, I.},
  title = {{Microscopic derivation of discrete hydrodynamics}},
  journal = {J. Chem. Phys.},
  year = {2009},
  volume = {131},
  pages = {244117}
}

@ARTICLE{LLNS_Espanol,
  author = {P. Espa{\~n}ol},
  title = {{Stochastic differential equations for non-linear hydrodynamics}},
  journal = {Physica A},
  year = {1998},
  volume = {248},
  pages = {77-96},
  number = {1-2}
}

@article{LLNS_FD_Fox,
  title={{Contributions to Non-Equilibrium Thermodynamics. I. Theory of Hydrodynamical Fluctuations}},
  author={Fox, R.F. and Uhlenbeck, G.E.},
  journal={Phys. Fluids},
  volume={13},
  pages={1893},
  year={1970}
}

@article{Boltzmann_FD_Fox,
  title={{Contributions to Nonequilibrium Thermodynamics. II. Fluctuation Theory for the Boltzmann Equation}},
  author={Fox, R.F. and Uhlenbeck, G.E.},
  journal={Phys. Fluids},
  volume={13},
  pages={2881},
  year={1970}
}

@ARTICLE{LLNS_Renormalization,
  author = {Forster, Dieter and Nelson, David R. and Stephen, Michael J.},
  title = {Large-distance and long-time properties of a randomly stirred fluid},
  journal = {Phys. Rev. A},
  year = {1977},
  volume = {16},
  pages = {732--749},
  number = {2},
}

@ARTICLE{Detonation_Fluctuations,
  author = {A. Lemarchand and B. Nowakowski},
  title = {Fluctuation-induced and Nonequilibrium-induced Bifurcations in a
	Thermochemical System},
  journal = {Mol. Simul.},
  year = {2004},
  volume = {30},
  pages = {773--780},
  number = {11-12}
}

@ARTICLE{GiantFluctuations_Nature,
  author = {Vailati, A. and Giglio, M.},
  title = {{Giant fluctuations in a free diffusion process}},
  journal = {Nature},
  year = {1997},
  volume = {390},
  pages = {262--265},
  number = {6657}
}

@ARTICLE{RayleighBernard_Fluctuations,
  author = {M. Wu and G. Ahlers and D.S. Cannell},
  title = {Thermally Induced Fluctuations below the Onset of {Rayleigh}-{B\'enard}
	Convection},
  journal = {Phys. Rev. Lett.},
  year = {1995},
  volume = {75},
  pages = {1743--1746},
  number = {9},
  numpages = {3}
}

\end{document}